\documentclass[%
 reprint,
superscriptaddress,
 amsmath,amssymb,
 aps,
]{revtex4-1}

\usepackage{graphicx}% Include figure files
\usepackage{dcolumn}% Align table columns on decimal point
\usepackage{bm}% bold math
\begin{document}

\preprint{APS/123-QED}

\title{\boldmath{$H$}-\boldmath{$T$} phase diagram of solid oxygen}% Force line breaks with \\

\author{T. Nomura}
%\altaffiliation[Also at ]
%{ISSP, University of Tokyo.}%Lines break automatically or can be forced with \\
\email{t.nomura@hzdr.de}
\affiliation{%
Institute for Solid State Physics, University of Tokyo, Kashiwa, Chiba 277-8581, Japan 
}%
\affiliation{%
Hochfeld-Magnetlabor Dresden (HLD-EMFL), Helmholtz-Zentrum Dresden-Rossendorf, D-01314 Dresden, Germany
}
%Dresden High Magnetic Field Laboratory (HLD-EMFL), Helmholtz-Zentrum Dresden-Rossendorf, 01328 Dresden, Germany
%Dresden High Magnetic Field Laboratory (HLD-EMFL), Helmholtz-Zentrum, Dresden-Rossendorf, D-01314 Dresden, Germany
%Hochfeld-Magnetlabor Dresden (HLD-EMFL), Helmholtz-Zentrum Dresden-Rossendorf, D-01314 Dresden, Germany
%Hochfeld-Magnetlabor Dresden (HLD-EMFL), Helmholtz-Zentrum Dresden-Rossendorf, D-01314 Dresden, Germany
%Hochfeld-Magnetlabor Dresden (HLD), Helmholtz-Zentrum Dresden-Rossendorf, Dresden D-01314, Germany
%Hochfeld-Magnetlabor Dresden (HLD), Helmholtz-Zentrum Dresden-Rossendorf, D-01314 Dresden, Germany

\author{Y. H. Matsuda}%
\email{ymatsuda@issp.u-tokyo.ac.jp} 
\affiliation{%
Institute for Solid State Physics, University of Tokyo, Kashiwa, Chiba 277-8581, Japan 
}%

\author{T. C. Kobayashi}
\affiliation{%
 Department of Physics, Okayama University, Okayama 700-8530, Japan}%

\date{\today}% It is always \today, today,

\begin{abstract}
Comprehensive magnetic-field-temperature ($H$-$T$) phase diagram of solid oxygen including the $\theta$ phase is discussed in the context of the ultrahigh-field measurement and the magnetocaloric effect (MCE) measurement.
The problems originating from the short duration of the pulse field, non-equilibrium condition and MCEs, are pointed out and dealt with.
The obtained phase diagram manifests the entropy relation between the phases as $S_\theta \sim S_\alpha<S_\beta<<S_\gamma$.

%\begin{description}
%\item[PACS numbers]75.30.Kz, 75.50.Ee, 75.50.Xx
%\end{description}
\end{abstract}

%\pacs{75.30.Kz, 75.50.Ee, 75.50.Xx}% PACS, the Physics and Astronomy% Classification Scheme.%\keywords{Suggested keywords}%Use showkeys class option if keyword%display desired
\maketitle

%%%%%%%%%%%
\section{introduction}
In the family of elemental solids, solid oxygen shows characteristic properties due to the magnetic moment of O$_2$ \cite{81PRB_DeFotis,85JPSJ_Uyeda,04PR_Freiman}.
The exchange interaction between O$_2$ molecules contributes to condensation energy in addition to the van der Waals interaction.
The strength of the exchange interaction greatly depends on the alignment of O$_2$ molecules \cite{83PRL_Hemert,93JCP_Bussery,08PCCP_Bartolomei,13JPSJ_Obata}, indicating that the packing structure of solid oxygen is tuned by its magnetic ground state.
Moreover, solid oxygen is the only antiferromagnetic (AFM) insulator composed of single element \cite{02LTP_Freiman}.
Thus, the phase diagram of solid oxygen is a unique playground for testing magnetism in an elemental solid.

In ambient pressure, three phases of solid oxygen appear, $\gamma$ (54.4--43.8 K), $\beta$ (43.8--23.9 K), and $\alpha$ (23.9-- K) \cite{04PR_Freiman}.
The $\gamma$ is a cubic phase ($Pm\overline{3}n$) where the molecules are rotating at each site.
This plastic phase is characterized by large entropy originating from the molecular rotation.
The $\beta$ is a rhombohedral phase ($R\overline{3}m$) where the molecular axis is ordered along $c$ direction.
Noteworthy, the entropy difference between two phases ($S_{\beta\gamma}=2.04R$, $R$ is the gas constant) is larger than melting \cite{29JACS_Giauq}.
In the $\beta$ phase, long range AFM order is suppressed by the geometrical frustration \cite{86_PRB_Step}.
The $\alpha$ is a monoclinic phase ($C2/m$), where the long range AFM order realizes with lattice deformation.

Recently, ultrahigh-field phase of solid oxygen, $\theta$ phase was discovered at around 120 T by using the single-turn coil (STC) \cite{14PRL_Nomura,15PRB_Nomura}.
The mechanism of the $\alpha$-$\theta$ phase transition is explained in analogy with the O$_2$-O$_2$ dimer, as follows.
In zero field, the most stable alignment of the O$_2$-O$_2$ dimer is H geometry where two molecules align rectangular-parallel \cite{83PRL_Hemert,93JCP_Bussery,08PCCP_Bartolomei,13JPSJ_Obata}.
The stability of the H geometry is owing to the maximized $\pi$ orbital overlapping between the dimer, leading to larger AFM interaction.
However, when the magnetization of the dimer is saturated by an external field, AFM interaction is no longer favored and the molecules struggle to decrease it.
For instance, if the molecules align in X geometry (crossed), overlap integral becomes zero due to the orthogonal geometry \cite{83PRL_Hemert,93JCP_Bussery,08PCCP_Bartolomei,13JPSJ_Obata}.
Namely, the O$_2$-O$_2$ dimer can tune the exchange interaction by changing its alignment.
When the strong magnetic field is applied to solid oxygen, it is natural to change the packing structure for smaller AFM interaction.
Recently, first-principle calculation also supports that the $\theta$ phase should have a crystal structure where molecules are packed with the canted alignment (cubic, $Pa\overline{3}$) \cite{16Kasamatsu}.

The thermodynamical $H$-$T$ phase diagram of solid oxygen including the $\theta$ phase has not yet been clarified.
The biggest obstacle is the requirement of the ultrahigh magnetic field above 100 T.
Such a high field is generated only by using the destructive pulse technique \cite{03LTP_Miura}.
Because of the short duration of the field, the phase transition occurs with large hysteresis indicating non-equilibrium condition \cite{14PRL_Nomura,15PRB_Nomura}.
In addition, temperature of the sample could change by the magnetocaloric effect (MCE) during the adiabatic magnetization \cite{17PRB_MCE_Nom,16JPSJ_Nomura}.
To obtain the thermodynamical phase boundary, these problems have to be dealt with.

In this paper, the thermodynamical $H$-$T$ phase diagram of solid oxygen is summarized as a compilation of the experimental works; STC \cite{14PRL_Nomura,15PRB_Nomura,16JPSJ_Nomura} and adiabatic MCE \cite{17PRB_MCE_Nom}.
In Sect. II, three major problems, (i) non-equilibrium, (ii) irreversible MCE, and (iii) reversible MCE are explained with results.
In Sect. III, sweep-speed dependence of the transition field is classified as a guide for dealing with the problems.
In Sect. IV, the $H$-$T$ phase diagram is carefully constructed with avoiding the problems.
We discuss the entropy relation between the phases in terms of the obtained phase diagram.
In Sect. V, conclusion is stated.

\section{results of the field-induced phase transitions}

\subsection{$\alpha$-$\theta$ and $\beta$-$\theta$ transitions in STC}
\begin{figure}[bt]
\centering
\includegraphics[width=8.4cm]{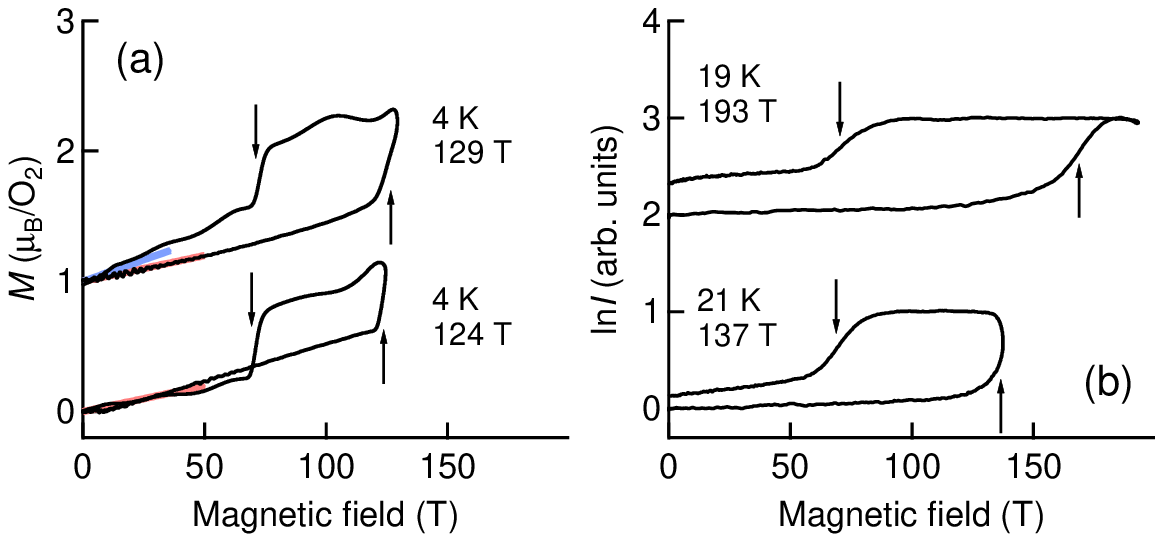}
\caption{\label{fig:Mag_Opt}
Results of the STC \cite{14PRL_Nomura,15PRB_Nomura}; (a) magnetization and (b) optical measurements.
The magnetization curves of the $\alpha$ (red) and $\beta$ (blue) phases are shown for comparison \cite{85JPSJ_Uyeda}.
The transmitted-light intensity is normalized for letting the change of ln$I$ be 1.
Upward (downward) arrows indicate $H_\mathrm{c}^+$ ($H_\mathrm{c}^-$).
Curves are shifted for clarity.
}
\end{figure}

The $\theta$ phase has been confirmed by the magnetization and magneto-optical measurements \cite{14PRL_Nomura,15PRB_Nomura}.
Representative results of the $\alpha$-$\theta$ phase transition are shown in Fig. \ref{fig:Mag_Opt}.
Initial temperature $T_0$, measured immediately before applying the field, and maximum field strength $H_\mathrm{max}$ are shown for each curve.
The phase transition was observed in the temperature range of $4<T_0<42$ K.
At the transition, the rapid increases of magnetization $M$ (Fig. \ref{fig:Mag_Opt}(a)) and transmitted-light intensity $I$ (Fig. \ref{fig:Mag_Opt}(b)) are observed.
The transition fields in up and down sweeps ($H_\mathrm{c}^+$, $H_\mathrm{c}^-$) are shown by the arrows.
As increasing $H_\mathrm{max}$, hysteresis becomes larger since the phase transition cannot follow the faster sweep of the pulsed field.
As a result, $H_\mathrm{c}^+$ and $H_\mathrm{c}^-$ are not uniquely defined even at the same $T_0$.
This is the problem (i), effect of non-equilibrium.
If we naively plot $H_\mathrm{c}^+$ and $H_\mathrm{c}^-$ as a function of $T_0$, the $H$-$T$ phase diagram is obtained with large deviation as Fig. \ref{fig:PDsum}(a).
In particular, the effect of non-equilibrium is serious in $H_\mathrm{c}^+$.

In Fig. \ref{fig:Mag_Opt}(a), the down sweep of the magnetization curve ($H_\mathrm{max}=129$ T, $T_0=4$ K) accords with the $\beta$ phase despite the initial phase is $\alpha$ \cite{85JPSJ_Uyeda}.
This is due to the irreversible heating during the $\alpha$-$\theta$ phase transition \cite{16JPSJ_Nomura}.
Dissipation related to the first-order phase transition (hysteresis loss and dissipative motion of domain walls) results in the heating effect.
Namely, $T$ differs from $T_0$ after the $\theta$ phase appears.
This is the problem (ii), irreversible MCE.
Fortunately, the irreversible MCE is relevant only when $H_\mathrm{max}$ is relatively higher since it is proportional to the amount of the induced $\theta$ phase.
The change of $T$ is not significant if the phase transition barely occurs like in the condition of $H_\mathrm{max}=124$ T where the down sweep of the magnetization accords with the $\alpha$ phase.

\subsection{$\beta$-$\gamma$ and $\alpha$-$\beta$ transitions observed in MCE}
\begin{figure}[bt]
\centering
\includegraphics[width=8.4cm]{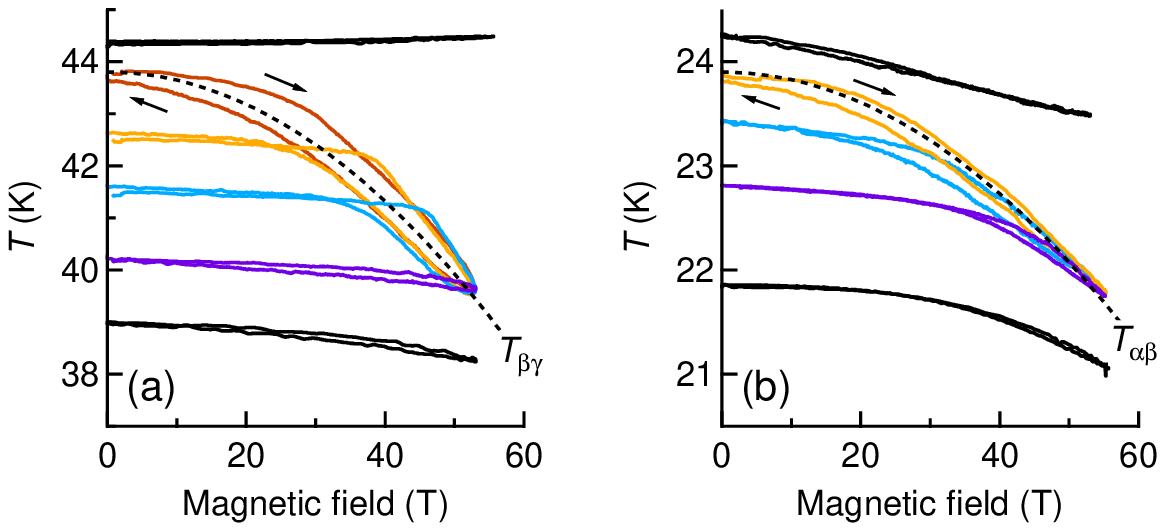}
\caption{\label{fig:MCEab}
MCE curves near the (a) $\beta$-$\gamma$ and (b) $\alpha$-$\beta$ phase boundaries \cite{17PRB_MCE_Nom}.
$H$ dependences of the phase boundaries are shown by dotted curves.
}
\end{figure}

The $\beta$-$\gamma$ and $\alpha$-$\beta$ transitions have been observed by the adiabatic MCE measurement \cite{17PRB_MCE_Nom}.
Figure \ref{fig:MCEab} shows the results near the (a) $\beta$-$\gamma$ and (b) $\alpha$-$\beta$ phase boundaries.
When these transitions occur in the adiabatic condition, $T$ changes along the phase boundary.
Here, entropy stays constant by balancing the $T$ and the fractions of the coexisting phases.
Since the measurement of the STC is adiabatic, $T$ changes from $T_0$ if the $\beta$-$\gamma$ and $\alpha$-$\beta$ phase transitions are induced.
This is the problem (iii), reversible MCE.
The reversible MCE is serious only beneath the $\beta$-$\gamma$ and $\alpha$-$\beta$ phase boundaries.
If no phase transition occurs, $\Delta T$ is less than 1 K even at 50 T \cite{17PRB_MCE_Nom}.

\section{Sweep-speed dependence of \boldmath{$H_\mathrm{c}$}}
\begin{figure}[tb]
\centering
\includegraphics[width=8.4cm]{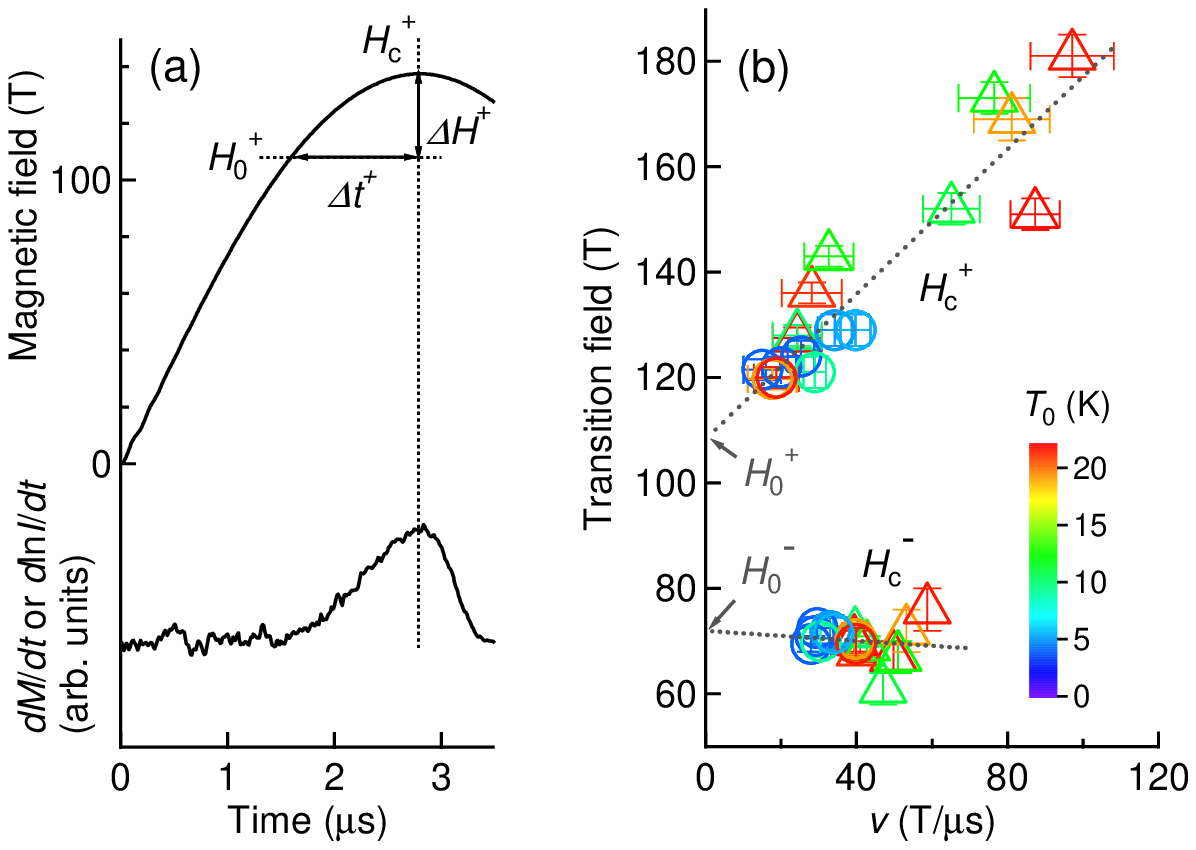}
\caption{\label{fig:v_dep}
(a) Graphical definition of the sweep speed $v=\Delta H/\Delta t$.
(b) $v$ dependence of the transition field \cite{15PRB_Nomura}. 
Circles and triangles show the results of the magnetization and optical measurements, respectively.
The color scale shows $T_0$ for each plot.
Dotted line shows the linear fitting of $H_\mathrm{c}$.
$H_\mathrm{0}$ is the expected $H_\mathrm{c}$ at $v=0$, and used for the definition of $v$ in the self-consistent way.
}
\end{figure}

We classify the sweep-speed $v$ dependence of the transition fields $H_\mathrm{c}$, as a guide for dealing with the problems.
In this paper, the superscript $+$ ($-$) is used for the up (down) sweep.
Figure \ref{fig:v_dep}(a) shows the definitions of $H_\mathrm{c}^+$ and $v^+$, which follow the way of Ref. \cite{15PRB_Nomura}.
$H_\mathrm{c}$ is defined by the extrema in $dM/dt$ or $d\mathrm{ln}I/dt$.
$v$ is self-consistently defined as the averaged sweep speed between the timings at $H=H_\mathrm{0}$ and $H_\mathrm{c}$.
$H_\mathrm{0}$ is the expected $H_\mathrm{c}$ in a quasi-static process ($v=0$ in Fig. \ref{fig:v_dep}(b)).
In this definition, $v$ is obtained as the sweep speed during the phase transition proceeds.

For the $\alpha$-$\theta$ phase transition, $H_\mathrm{c}$ is plotted as a function of $v$ in Fig. \ref{fig:v_dep}(b).
The data at $T_0 < 22$ K are plotted and equally treated since $H_\mathrm{c}$ seems to be independent on $T_0$ in this temperature range.
By the linear fitting, we obtained the $v$ dependence of $H_\mathrm{c}$ as
\begin{eqnarray}
H^+_\mathrm{c}&=&0.69(5) \ v^+ + 108(2) \ \mathrm{T},\\
  \label{scl1}
H^-_\mathrm{c}&=&-0.05(6) \ v^- + 72(2) \ \mathrm{T}.
  \label{scl2}
\end{eqnarray}

By taking the slope 0.69 for up sweep, we corrected $H_\mathrm{c}^+$ to the limit of $v=0$. 
The corrected transition field $H_\mathrm{c0}^+$ corresponds to the value expected at quasi-equilibrium condition.
For discussing the thermodynamical phase diagram, $H_\mathrm{c0}^+$ is more appropriate since the effect of non-equilibrium is suppressed.
In this study, we applied this correction for all temperature range assuming that the $\beta$-$\theta$ phase transition also follows the same $v$ dependence.
For the down sweep, the correction was not applied since $B_\mathrm{c}^-$ does not depend on $v^-$ greatly.

\section{phase diagram}

\begin{figure*}[ptb]
\centering
\includegraphics[width=15cm]{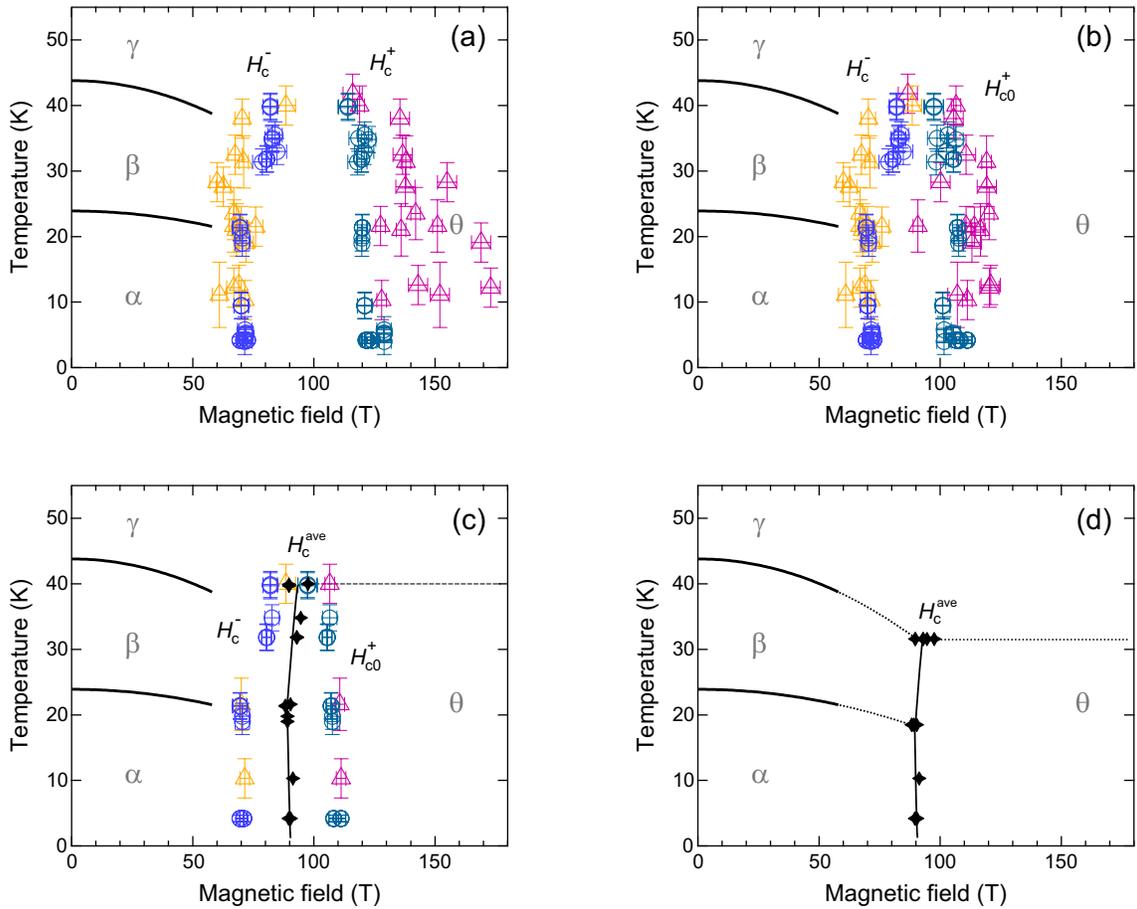}
\caption{\label{fig:PDsum}
$H$-$T$ phase diagram of solid oxygen.
(a) $H_\mathrm{c}^+$ and $H_\mathrm{c}^-$ are plotted as a function of $T_0$ without any modifications.
Circles and triangles show the results of the magnetization and optical measurements, respectively.
%The solid curves show the $\beta$-$\gamma$ and $\alpha$-$\beta$ phase boundaries \cite{17PRB_MCE_Nom}.
(b) $H_\mathrm{c0}^+$ is plotted instead of $H_\mathrm{c}^+$.
(c) The data plots obtained in slower sweep speed ($15<v^+<30$ T/$\mu$s) are selected.
The averaged transition field $H_\mathrm{c}^\mathrm{ave} = (H_\mathrm{c0}^+ + H_\mathrm{c}^-)/2$ is plotted by black star with the guiding curve.
The gray dashed line is the $\theta$-$\gamma$ phase boundary expected from the entropy relation.
(d) Proposed $H$-$T$ phase diagram of solid oxygen.
The reversible MCEs along the $\beta$-$\gamma$ and $\alpha$-$\beta$ phase boundaries are taken into account.
}
\end{figure*}

In this section, the $H$-$T$ phase diagram is corrected from Fig. \ref{fig:PDsum}(a) to \ref{fig:PDsum}(d) by dealing with the problems one by one.
Figure \ref{fig:PDsum}(a) is the naively obtained phase diagram, where $H_\mathrm{c}^+$ and $H_\mathrm{c}^-$ are plotted as a function of $T_0$.
The $\beta$-$\gamma$ and $\alpha$-$\beta$ phase boundaries obtained by the adiabatic MCE measurement are shown by solid curves \cite{17PRB_MCE_Nom}.
Deviation of $H_\mathrm{c}^+$ is due to the non-equilibrium of the $\alpha$-$\theta$ and $\beta$-$\theta$ phase transitions \cite{14PRL_Nomura,15PRB_Nomura}.
The $v$ dependence of $H_\mathrm{c}^+$ has to be taken into account for further discussions.

Figure \ref{fig:PDsum}(b) shows the corrected phase diagram plotting $H_\mathrm{c0}^+$ instead of  $H_\mathrm{c}^+$.
By the correction, the deviation originating from the problem (i) is greatly suppressed.
For discussing the thermodynamical phase diagram, it is usual to read the center of hysteresis as $H_\mathrm{c}^\mathrm{ave} = (H_\mathrm{c0}^+ + H_\mathrm{c}^-)/2$.
However, we should be careful for the temperature change after the field-induced phase transition.
Especially after the $\alpha$-$\theta$ phase transition, irreversible heating as much as 700 J/mol related to the hysteresis loss was observed \cite{16JPSJ_Nomura}.

To avoid the effect of irreversible MCE (problem (ii)), data plots obtained only with slower $v$ ($15<v^+<30$ T/$\mu$s) are selected in Fig. \ref{fig:PDsum}(c).
Slower $v$ implies that the phase transition barely occurs at the top of the field.
Irreversible MCE is considered to be proportional to the fraction of the field-induced $\theta$ phase.
Therefore, the temperature difference between the up and down sweeps are greatly reduced in this condition.
The averaged transition field $H_\mathrm{c}^\mathrm{ave}$ is plotted by black star.
The Clausius-Clapeyron equation
\begin{equation}
dT_\mathrm{c}/dH_\mathrm{c} =-\Delta M/\Delta S
\label{eq:cc_eq}
\end{equation}
suggests that the obtained steep phase boundary means large $\Delta M$ and small $\Delta S$ \cite{15PRB_Nomura}.

Here, we quantitatively discuss the entropy relation between the $\alpha$, $\beta$, $\gamma$, and $\theta$ phases.
The slope of the phase boundaries are roughly estimated as $5<|dT/dB|_{\alpha\theta}$, $1<(dT/dB)_{\beta\theta}<10$ in the units of K/T.
The differences of the magnetization are roughly estimated as $\Delta M_{\alpha\theta}=1.2\mu_\mathrm{B}$/O$_2$ and $\Delta M_{\beta\theta}=1\mu_\mathrm{B}$/O$_2$ \cite{15PRB_Nomura}.
Here, we assumed that the magnetization of the $\theta$ phase is saturated ($M_\theta = 2\mu_\mathrm{B}$/O$_2$).
Based on Eq. (\ref{eq:cc_eq}), the entropy differences are obtained as $|\Delta S|_{\alpha\theta}<0.16R$ and $-0.67R<\Delta S_{\beta\theta}<-0.07R$.
At zero field, the entropy relation between the $\alpha$, $\beta$, and $\gamma$ phases are already known as $S_\alpha<S_\beta<<S_\gamma$ with the differences of $\Delta S_{\alpha\beta}=0.47R$ and $\Delta S_{\beta\gamma}=2.04R$ \cite{29JACS_Giauq}.
Summarizing them, the entropy relation is obtained as $S_\theta \sim S_\alpha<S_\beta<<S_\gamma$.
In the following, we assume the $\gamma$ phase survives at high fields \cite{g_comment} and discuss how the four phases connect each other in the $H$-$T$ phase diagram.

Large $\Delta S_{\theta\gamma}$ suggests that the $\gamma$-$\theta$ phase boundary is flat.
In addition, $\Delta M_{\theta\gamma}$ would be almost zero since the magnetizations of the $\theta$ and $\gamma$ phases almost saturate at 150 T.
Therefore, the $\theta$-$\gamma$ phase boundary has to be completely flat as shown by the gray dashed line in Fig. \ref{fig:PDsum}(c).
Here, the $\beta$-$\gamma$-$\theta$ triple point is not clear since the extrapolated $\beta$-$\gamma$ phase boundary does not smoothly connect to the $\theta$-$\gamma$ boundary.

\begin{figure}[tb]
\centering
\includegraphics[width=8.2cm]{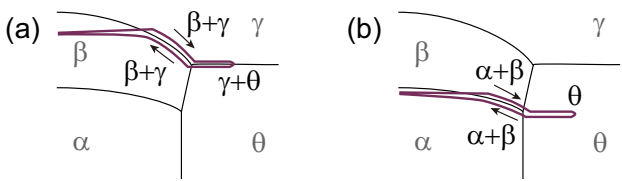}
\caption{\label{fig:MCE_sc}
Expected MCE curves beneath the (a) $\beta$-$\gamma$ and (b) $\alpha$-$\beta$ phase boundaries.
}
\end{figure}

For explaining the $\beta$-$\gamma$-$\theta$ triple point, the reversible MCE (problem (iii)) has to be taken into account.
The schematic MCE curve is shown in Fig. \ref{fig:MCE_sc}(a).
When the magnetic field passes the $\beta$-$\gamma$ phase boundary in the adiabatic condition, $T$ decreases along the phase boundary with increasing the fraction of the $\gamma$ phase.
For example, in the case of $T_0 = 42$ K, 100 T is necessary for the entire phase transition from the $\beta$ to $\gamma$ phase \cite{17PRB_MCE_Nom}.
That means the plots beneath the $\beta$-$\gamma$ phase boundary are cooled down to the $\beta$-$\gamma$-$\theta$ triple point with phase coexistence.
Finally at the $\beta$-$\gamma$-$\theta$ triple point, the $\beta$ phase will transform to the $\theta$ phase, which is detected in the experiment.
At higher field than the triple point, the MCE curve will follow the $\gamma$-$\theta$ phase boundary with phase coexistence.
However, this phase coexistence cannot be confirmed by the magnetization and optical spectroscopy since the $\gamma$ and $\theta$ phases show similar results in these measurements.

The same behavior of MCE would occur also for the $\alpha$-$\beta$ phase boundary.
Therefore, the data plots around 20 K would also reach to the $\alpha$-$\beta$-$\theta$ triple point as shown in Fig. \ref{fig:MCE_sc}(b).
At the triple point, the $\alpha$ and $\beta$ phases wholly transform to the $\theta$ phase.
When the $\beta$-$\gamma$ and $\alpha$-$\beta$ phase transitions are not involved ($T_0 <15$ K), the effect of the reversible MCE is less than 1 K and negligible \cite{17PRB_MCE_Nom}.

Finally, we propose the $H$-$T$ phase diagram of solid oxygen as Fig. \ref{fig:PDsum} (d).
Here, we assume that the data plots around 40 K and 20 K get together at the triple points.
The $\theta$-$\gamma$ transition (solid-plastic transition) is expected to occur at 31 K.
This is close to the value of the solid-plastic transition of N$_2$ ($\alpha$-$\beta$ transition, $T_{\alpha\beta}=35.6$ K) \cite{09LTP_Grzebyk}.

The thermodynamical $\alpha$-$\theta$ transition field is obtained at around 90 T.
Recently, such an ultrahigh magnetic field becomes accessible using non-destructive pulse magnets and applied to the researches in condensed matter physics \cite{12PNAS_Jaime,12PRL_Altarawneh,15PRL_Kim,15PRBR_Zherlitsyn}.
If the magnetic field is generated for few ms, $H_\mathrm{max}=100$ T might be enough to observe the $\theta$ phase since hysteresis becomes smaller.
In this time scale, various kinds of measurement such as magnetostriction, ultrasound, and x-ray diffraction are applicable to obtain the structural information of the $\theta$ phase.
These measurements are important for revealing the whole picture of the $\theta$ phase and for further understanding of oxygen.

\section{conclusion}
The thermodynamical $H$-$T$ phase diagram of solid oxygen including the $\theta$ phase was proposed as Fig. \ref{fig:PDsum}(d).
The phase diagram was obtained by analyzing the results of the magnetization, optical, and adiabatic MCE measurements with avoiding the problems originating from the short duration of the field.
Using the Clausius-Clapeyron equation, the entropy relation between the phases was obtained as $S_\theta \sim S_\alpha<S_\beta<<S_\gamma$.

The corrections of the phase diagram applied in Figs. \ref{fig:PDsum} are abstracted as follows.
From (a) to (b), $H_\mathrm{c}^+$ was corrected to $H_\mathrm{c0}^+$ by using the $v$ dependence of $H_\mathrm{c}^+$ for suppressing the effect of (i) non-equilibrium.
From (b) to (c), only the data obtained in slower $v$ were collected for reducing the effect of (ii) irreversible MCE.
Here, $H_\mathrm{c}^\mathrm{ave}$ was introduced for discussing the thermodynamical phase boundary.
From (c) to (d), the effect of (iii) reversible MCE was taken into account for the consistency of the $\beta$-$\gamma$-$\theta$ triple point.
In the adiabatic condition of the STC, $T$ is expected to decrease along the $\beta$-$\gamma$ and $\alpha$-$\beta$ phase boundaries and to reach to the triple points.

The ground state of oxygen obviously depends on the external magnetic field since O$_2$ is a magnetic molecule.
The obtained $H$-$T$ phase diagram will contribute for discussing the thermodynamical stability of oxygen. 
Potentially, it could lead to controlling the chemical activity of oxygen by magnetic field, during the biological reaction or material processing \cite{15PRL_Kurahashi,01PhysicaB_Kita}.

\section*{acknowledgments}
The authors acknowledge S. Takeyama, K. Kindo, A. Matsuo, Y. Kohama, and A. Ikeda for fruitful discussions.
T. N. was supported by Japan Society for the Promotion of Science through the Program for Leading Graduate Schools (MERIT), a Grant-in-Aid for JSPS Fellows, and HLD at HZDR, member of the European Magnetic Field Laboratory.
This work was partly supported by JSPS KAKENHI, Grant-in-Aid for Scientific Research (B) (16H04009).


\begin{thebibliography}{99}
\bibitem{g_comment}
The results of the $\gamma$ phase at above 100 T were not reproducible.
Most probably, it is because of the plasticity; the crystal cracks and moves during the pulsed field.
\end{thebibliography}

\end{document}